\documentclass[12pt,aps,prd,superscriptaddress,nofootinbib,showpacs,preprintnumbers]{revtex4}
\usepackage{amsfonts,amssymb,amsmath}
\usepackage{multirow}
\usepackage{color}
\usepackage{graphicx,epsfig,rotating}
\usepackage{pstricks}
\usepackage{slashed}

\usepackage[normalem]{ulem}

\definecolor{darkgreen}{cmyk}{1,0,1,0.6}

\newcommand{\eea}{\end{eqnarray}}
\newcommand{\gev}{\,{\rm GeV}}
\newcommand{\tev}{\,{\rm TeV}}

\newcommand{\ba}{\begin{array}}
\newcommand{\ea}{\end{array}}
\newcommand{\beq}{\begin{equation}}
\newcommand{\eeq}{\end{equation}}

\def\bt{\begin{table}}
\def\et{\end{table}}
\def\bc{\begin{center}}
\def\ec{\end{center}}
\def\bi{\begin{itemize}}
\def\ei{\end{itemize}}
\def\bea{\begin{eqnarray}}
\def\eea{\end{eqnarray}}
\def\beas{\begin{eqnarray*}}
\def\eeas{\end{eqnarray*}}

\def\dis{\displaystyle}
\def\bnq{\begin{eqnarray}}
\def\enq{\end{eqnarray}}
\def\barr{\begin{array}}
\def\earr{\end{array}}
\def\gev{\, {\rm GeV}}

\def\lapp{\mathrel{\rlap{\raise.5ex\hbox{$<$}}
                    {\lower.5ex\hbox{$\sim$}}}}
\def\gapp{\mathrel{\rlap{\raise.5ex\hbox{$>$}}
                    {\lower.5ex\hbox{$\sim$}}}}

\begin{document} 

\setlength{\parskip}{0.1cm}

\title{\sc Dijet signals of the Little Higgs Model with T-parity}

\author{Debajyoti Choudhury}    \email{debchou@physics.du.ac.in}
\affiliation{Department of Physics and Astrophysics, 
University of Delhi, Delhi 110007, India.}

\author{Dilip Kumar Ghosh}\email{tpdkg@iacs.res.in}
\affiliation{Department of Theoretical Physics, 
Indian Association for the Cultivation of Science, 
2A \& 2B Raja S.C. Mullick Road, Kolkata 700 032, India.}
 
\author{Santosh~Kumar~Rai}\email{santosh.rai@okstate.edu}
\affiliation{Department of Physics and Oklahoma Center for High
Energy Physics, Oklahoma State University, Stillwater, OK
74078-3072, USA}

\vskip 30pt
\begin{abstract}
The Littest Higgs model with $T$-parity (LHT), 
apart from offering a viable solution to the
naturalness problem of the Standard Model, 
also predicts a set of new fermions as well as a candidate for
dark matter. 
We explore the possibility of discovering the heavy $T$-odd quark $Q_H$
at the LHC in a final state comprising two hard jets with a 
large missing transverse momentum.
Also discussed is the role of heavy flavor tagging.
\end{abstract}



\preprint{OSU-HEP-12-03}

\maketitle

\vfill
\section{Introduction}
\label{sect:Intro}
Little Higgs models \cite{lh0} offer an intriguing resolution of the
fine-tuning problem associated with electroweak symmetry breaking.
Incorporating the standard model (SM) Higgs as a pseudo-Goldstone
boson of some global symmetry which is spontaneously broken at a scale
$\Lambda ( \equiv 4\pi f) \sim 10 $ TeV, the low energy effective
theory is described by a non-linear sigma model.  With the
introduction of new gauge bosons and partners of the top quark with
masses of the order of $f$, the quadratically divergent contributions
to the Higgs mass are exactly cancelled at one loop level, thereby
ameliorating the fine-tuning problem.

However, constraints from precision electroweak measurements imply
that the scale $f$ needs to be above $\sim 5$ TeV \cite{lh_ew}. For
such a large value of $f$, one faces the re-introduction of a fine
tuning between the cutoff scale ($\sim 4\pi f$) for the model and the
weak scale.  To circumvent this serious problem of the original Little
Higgs model, a new discrete symmetry, called $T$-parity (and analogous
to the $R$ parity in the minimal supersymmetric standard model), was
introduced. The Littlest Higgs Model with $T$-parity (LHT)
\cite{lht1,lht2,lht3,lht4} provides a fully realistic and consistent
model which satisfies the electroweak precision data. 
All SM fields are $T$-even under this new symmetry, while
the new heavy partners are $T$-odd, and
 can only be produced 
in pairs. Moreover, even after electroweak 
symmetry breaking,  
mixing between the SM gauge bosons and their $T$-odd 
counterparts is prohibited, thereby removing any tree level new physics 
contribution to the electroweak precision observables. Consequently,
all new physics corrections now appear only at the
one loop level or higher, and, hence, are naturally small.
As a result, 
the EW precision data concede a relatively low value 
of the new particle mass scale $f\sim 500$ GeV \cite{lht3}, 
thereby allowing copious 
production of different $T$-odd heavy partners of the SM particles at the 
LHC and future $e^+ e^-$ linear collider
(ILC)~\cite{lht2,wyler,Belyaev:2006jh,Carena:2006jx,
Chen:2006ie,Choudhury:2006mp,Cacciapaglia:2009cu,Cacciapaglia:2009cv}.

A further interesting feature of 
$T$-parity is the prediction of a colorless neutral weakly 
interacting stable $T$-odd particle (LTP) $A_H$, the heavy 
partner of the hypercharge gauge boson; known as the 
{\it heavy photon}, it is a good candidate for cold dark 
matter \cite{lht_dm}.

In this paper, we revisit the LHC signatures of $T$-odd
heavy quark pair production within this model. As with a host of other
models for new physics beyond the SM, signatures in a hadronic
environment are often the easiest to tag on to when cascade decays
(hopefully, with isolated hard leptons) are considered \cite{wyler,
  Belyaev:2006jh,Carena:2006jx,Choudhury:2006mp,
  Cacciapaglia:2009cu,Bhattacherjee:2009jh}.  This, indeed, happens in
the LHT models for a significant range of parameters. However, for a
large range, cascade chains do not occur and the $T$-odd quarks decay
promptly into the $A_H$ and a SM quark. The consequent final state,
namely a dijet pair alongwith missing transverse energy, is relatively
more difficult to analyse and this had led to search strategies
ignoring this important part of the parameter space.  In the case of
the third generation (down-type) heavy $T$-odd quark pair production,
the final state jets, when tagged, give rise to $2b$-jet final state,
while untagged jets contribute to the dijet cross-section from the
pair production of first two generation $T$-odd heavy quarks.
Performing a detailed estimation of the observability of this signal,
and taking into account all relevant SM backgrounds, we delineate the
additional part of the LHT parameter space that is amenable to
discovery at the LHC.  

The rest of the paper is organized as
follows. In Section \ref{sect:model}, we briefly discuss the main
features of the model. In Section \ref{sect:t-oddprod}, we discuss
pair production of $T$-odd heavy quarks and their 
 two body decay into standard model quarks and the LTP, $A_H$. In
Section \ref{sect:result_analysis}, after discussing signal and
background events, we estimate the detectability of the LHT signal in
the dijet plus missing energy channel at the LHC. Finally, our conclusions
are given in Section \ref{sect:concls}.

\section{The Model} 
\label{sect:model} 
Rather than attempting a detailed study of the Littlest Higgs model
with $T$-parity (see, for example, Refs.\cite{lht1,lht2,lht3}), we 
concentrate on the issues directly relevant to us.
Considering a non-linear sigma 
model with a $SU(5)$ global symmetry, let us gauge the 
subroup $[SU(2) \times U(1)]_1 \times [SU(2) \times U(1)]_2$. 
A discrete symmetry ($T$-parity)
exchanges the two $[SU(2) \times U(1)]$ units, thereby restricting the
matter content as well as the gauge couplings.  The global 
$SU(5)$ is broken down to 
$SO(5)$ at some high scale $f$, 
leading to 14 massless Nambu-Goldstone (NG) bosons \cite {lh0}. 
Simultaneously, the gauged symmetry 
is broken down to the diagonal
subgroup $SU(2)_L\times U(1)_{\rm Y} $ to be 
identified with the SM gauge
group. Of the 14 NG bosons, four are manifested as the 
longitudinal modes of the 
heavy gauge bosons. The remaining ten 
decompose into a $T$-even $SU(2)$ doublet $h$, identified with
the SM Higgs field, and a complex $T$-odd $SU(2)$ triplet $\Phi $,
which obtains a mass 
$M_{\Phi} = \sqrt{2} M_{h} f/v_{\rm SM}$ at one
loop, with $M_h$ being the SM Higgs mass. 

Not being singlets under the SM gauge
group, the  $T$-odd acquire further 
contributions to their masses from electroweak symmetry breaking, 
and we have 
\beq
\barr{rclcl}
& & M_{A_H} & \simeq &  \dis 
\frac{g^\prime f}{\sqrt{5}}\left[1 - \frac{5v^2_{\rm SM}}{8 f^2}
+...\right] \ , 
\\[2ex]
M_{Z_H} & \simeq &  M_{W_H} & = & \dis g f \left[ 1 - \frac{v^2_{\rm SM}}{8 f^2}+...\right] \ .
\earr
   \label{eq:gauge_mass}
\eeq
Here, $v_{\rm SM}\simeq 246 $ GeV is the electroweak symmetry breaking
scale. Since $g^\prime < g $, $A_H$ is substantially lighter than 
other two $T$-odd heavy gauge bosons $W_H $ and $Z_H$. 

Consistent implementation of $T$-parity in the 
fermion sector requires that each SM fermion doublet must be replaced 
by a pair of fields $F_i (i =1,2)$ \cite{lht1,lht2,lht3}, 
where each $F_i$ is a
doublet under $SU(2)_i$ and singlet under the other.  
Under $T$-parity, $F_1 \leftrightarrow F_2 $ and 
the $T$-even
combination of $F_i$ is identified with the SM fermion doublet.
The other ($T$-odd) combination is its heavy partner $(F_H)$. To
generate mass terms for the latter, one requires 
an extra set of $T$-odd $SU(2)$ singlet fermions
in the theory~\cite{lht1,lht2,lht3}. Considering an
universal and flavour
diagonal Yukawa coupling $\kappa $ for $U_H $ and $D_H$ (the $T$-odd
heavy partners of the SM quarks $(u,c)$ and $(d,s)$ respectively), we
have
\bnq
M_{D_H} \simeq \sqrt{2} \, \kappa \, f \ ,
\qquad
M_{U_H}  \simeq \sqrt{2} \, \kappa \, f \, 
         \left(1 - \, \frac{v^2_{\rm SM}}{8 \, f^2} \right) \ .
\label{eq:Toddmass}
\enq 
Since $f \gapp 500 \gev$, it is evident
from eq.(\ref{eq:Toddmass}) that
the up-- and down--type $T$-odd heavy partners have nearly equal
masses. In summary, the complete
spectrum of the LHT model with $T$-parity relevant for our analysis
will only depend on two free parameters: the new physics scale $f$ and
the flavour independent Yukawa coupling $\kappa$.

\section{Production and decay of the $1^{\rm st}$ and 
$2^{\rm nd}$ generation $T$-odd heavy quark}
\label{sect:t-oddprod}
Given the model described in the preceding section, we may calculate
the production rates of $T$-odd quarks at the LHC.  The latter can be
copiously pair produced ($Q_H \bar Q_H$) as long as their masses are
not too large.  The LHC being primarily a gluon machine, the pure QCD
process naturally dominates, and the calculation thereof is identical
to that for any heavy quark \cite{Combridge:1978kx}. However, even the
electroweak amplitudes do have substantial contributions, especially
for like-sign $Q_H$ production \cite{Cacciapaglia:2009cu}.  
To discuss these, we need to know the electroweak 
couplings of the $T$-odd quarks. 

The $Q_H q^{(\prime)} V_H$ couplings (where $V_H$ is one of $W_H$, $Z_H$ and 
$A_H$) 
depend on $f$.  The couplings $U_H-d-W_H$ and $D_H-u-W_H$ are of equal
strength owing to $SU(2)$ invariance of the Lagrangian, viz.
\beq
     g_{U_H d W_H} = g_{D_H u W_H} = g / \sqrt{2} \ .
\eeq
On the other hand,
the couplings to the $Z_H$ and $A_H$ have a crucial dependence 
on isospin ($T_3$), namely
\beq
g_{f_H f Z_H} = g \, c_H \, T_{3f} + g' \, s_H \, Y' \ ,
\qquad
g_{f_H f A_H} = - g \, s_H \, T_{3f} + g' \, c_H \, Y' \ ,
\label{f_ah_coup}
\eeq
where $Y' = - 1/ 10$ and $\theta_H$ is the Weinberg angle in the heavy sector, 
viz. 
\beq
  s_H  \equiv \sin\theta_H \simeq
 \frac{5 \, g \, g'} {5 \, g^2 - g'^2} \; \frac{v_{\rm SM}^2}{4 \, f^2}
\ ,
\qquad
  c_H  \equiv \cos\theta_H \ .
\eeq
Eq.(\ref{f_ah_coup}) immediately opens up the possibility for a 
cancellation in
$g_{D_H d A_H}$,  especially for  smaller $f$ values. 

Before discussing the production, let us comment about its aftermath,
namely the decays.  Once these heavy $T$-odd quarks are produced, they
will promptly decay into ($T$-even) SM quarks and $T$-odd heavy gauge
bosons $(W^\pm_H, Z_H,A_H)$.  As the $W^\pm_H/ Z_H$ are themselves
unstable, decays into these channels would lead to cascades and the
corresponding signatures have been well-studied in the literature,
albeit for differing regions in the parameter space.  Instead, we
concentrate on the decays $Q_H \to q + A_H$.  As a comparison of
eq.(\ref{eq:Toddmass}) with eqs.(\ref{eq:gauge_mass}) shows, the $U_H$
and $D_H$ are heavier than $W_H/Z_H$ only if $\kappa$ is not too 
small\footnote{Note that a very small $\kappa$ is
disallowed as this would render $Q_H$ to be stable and, hence,
lead to a colored and charged dark matter candidate! On the
other hand, in the fine-tuned case of $Q_H$ being only
marginally heavier than the $A_H$, it would be quasi-stable on
detector scales, leading to the formation of states analogous to
 $R$-hadrons~\cite{ArkaniHamed:2004yi}. We shall not consider such finely-tuned scenarios.}.  
Also important are the parameter-dependences of the $f_H f' V_H$ couplings, in particular
the suppression of $\Gamma(D_H \to d + A_H)$ for small $f$.

\begin{figure}[!t]
\begin{center}
\includegraphics[width=3.6in]{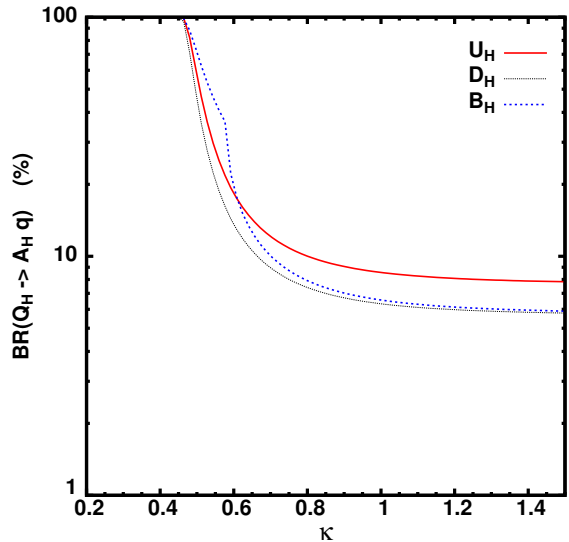}
\end{center}
\vspace*{-0.9cm}
\caption{\em Variation of the branching ratio of heavy quarks ($Q_H$)
into $q + A_H$ in the LHT model with the  parameter $\kappa$ for a 
fixed value of the scale parameter $f=1000$ GeV.}
\label{fig:brfig}
\end{figure}

In Fig.~\ref{fig:brfig}, we display the two-body decay branching
probability for the T-odd quarks $U_H, D_H $ and $ B_H $.  For $\kappa
\lapp 0.45$, the two-body decay mode into $A_H$ is the only
kinematically allowed one\footnote{For $T_H$, depending on the value
of $f$, it could even be that $T_H \to A_H + b + W^+$ is the only
one allowed.}.  For $\kappa \gapp 0.45$, though, the other modes are
accessible, and the two-body branching into $A_H$ drops very fast,
essentially on account of the larger coupling to the $W_H$.  For a sufficiently 
large $\kappa$, the kinematical suppression is rendered irrelevant and the 
branching fraction is determined only in terms of the coupling constants.  
The branching for the $B_H $ shows a small kink at $\kappa \sim 0.6$ as the 
$t W_H$ decay mode becomes available only at this juncture\footnote{Clearly, the 
exact location of this kink depends on the value of $f$.}.

We now turn to the production of $Q_H$ pairs.  As the heavy quarks
corresponding to the first two generations are nearly degenerate, and
lead to very similar final state configurations, we sum over all four
flavours. Understandably, the pure-QCD processes $g g \to Q_{iH} \bar
Q_{iH}$ and $q_j \bar q_j \to Q_{iH} \bar Q_{iH}$ tend to dominate and
was considered in Ref~\cite{Perelstein:2011ds}. The ordinary
electroweak contributions to the $q \bar{q}$-initiated process is only
${\cal O}(\alpha_{\rm wk}^2)$.  As for the $V_H$ mediated
contributions, these too are only ${\cal O}(\alpha_{\rm wk}^2)$ unless
$i = j$, whence it can be ${\cal O}(\alpha_s \,\alpha_{\rm wk})$.
Note, though, that $\kappa \lapp 0.45$ means that the $W_H/Z_H$ are at
least as massive as the $Q_H$ and this implies additional suppression;
and while the $A_H$-diagram is not suppressed kinematically, it is
virtually irrelevant for $D_H$ production (owing to the smaller
cross-sections). Thus, in effect, most of the electroweak
contributions are expected to be rather subdominant.  However, we must
note that for large values of $f$, the partons are required to carry a bigger 
momentum fraction $x$ of the proton. 
This, in turn, renders the valence quark induced subprocesses (mediated by $V_H$)  
to be comparable to or even dominate the gluon initiated subprocesses. Thus, processes 
such as $u \, q_j  \to U_{H} Q_{jH}$  (where j=1,2) cannot be neglected anymore as, for 
example, was done in Ref~\cite{Perelstein:2011ds}. 
We, on the other hand, include all processes (and all amplitudes) that lead to the production 
of a pair of heavy  quarks ($Q_{iH} Q_{jH} \, (\bar Q_{jH})$), irrespective of the
flavour composition.

\begin{figure}[!h]
\begin{center}
\includegraphics[width=3.2in]{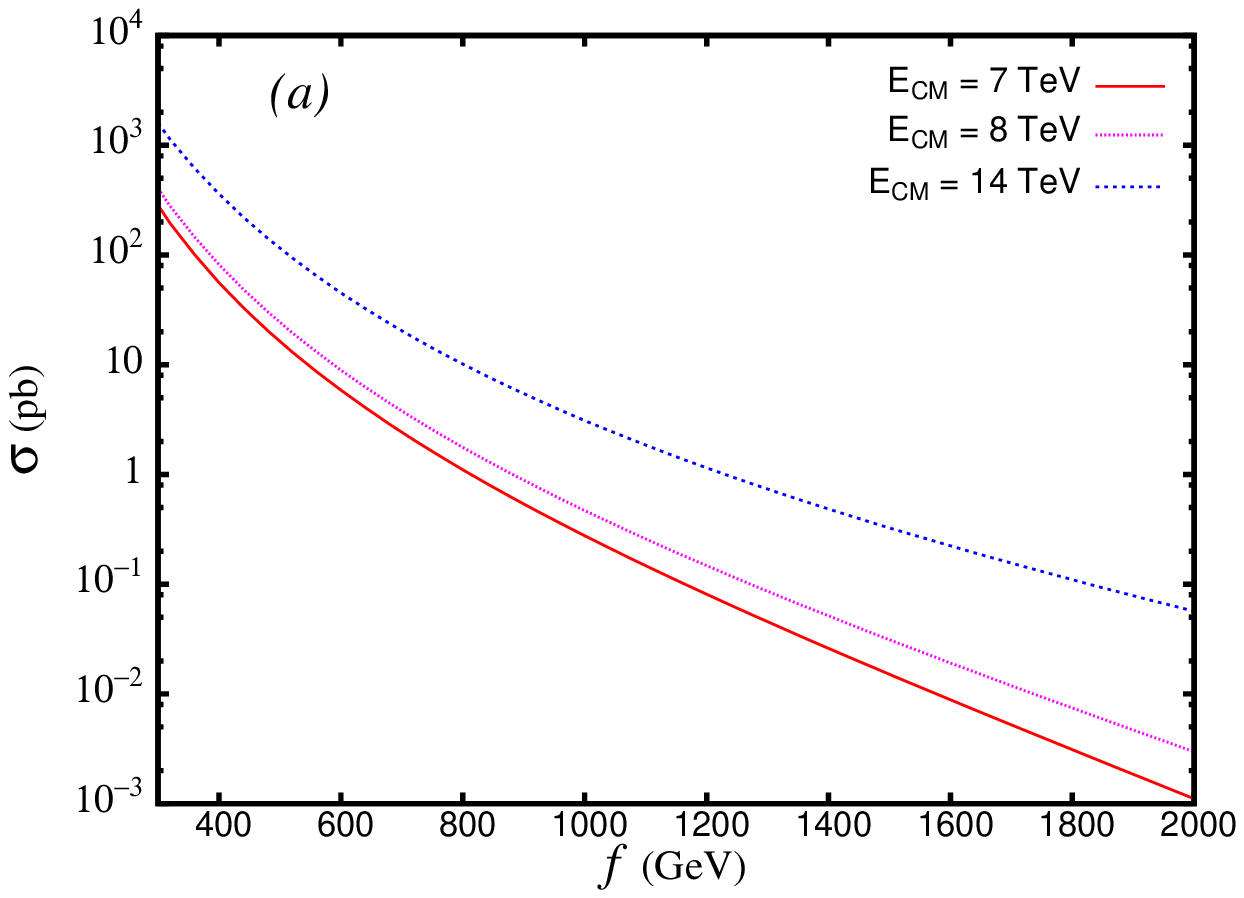}
\includegraphics[width=3.2in]{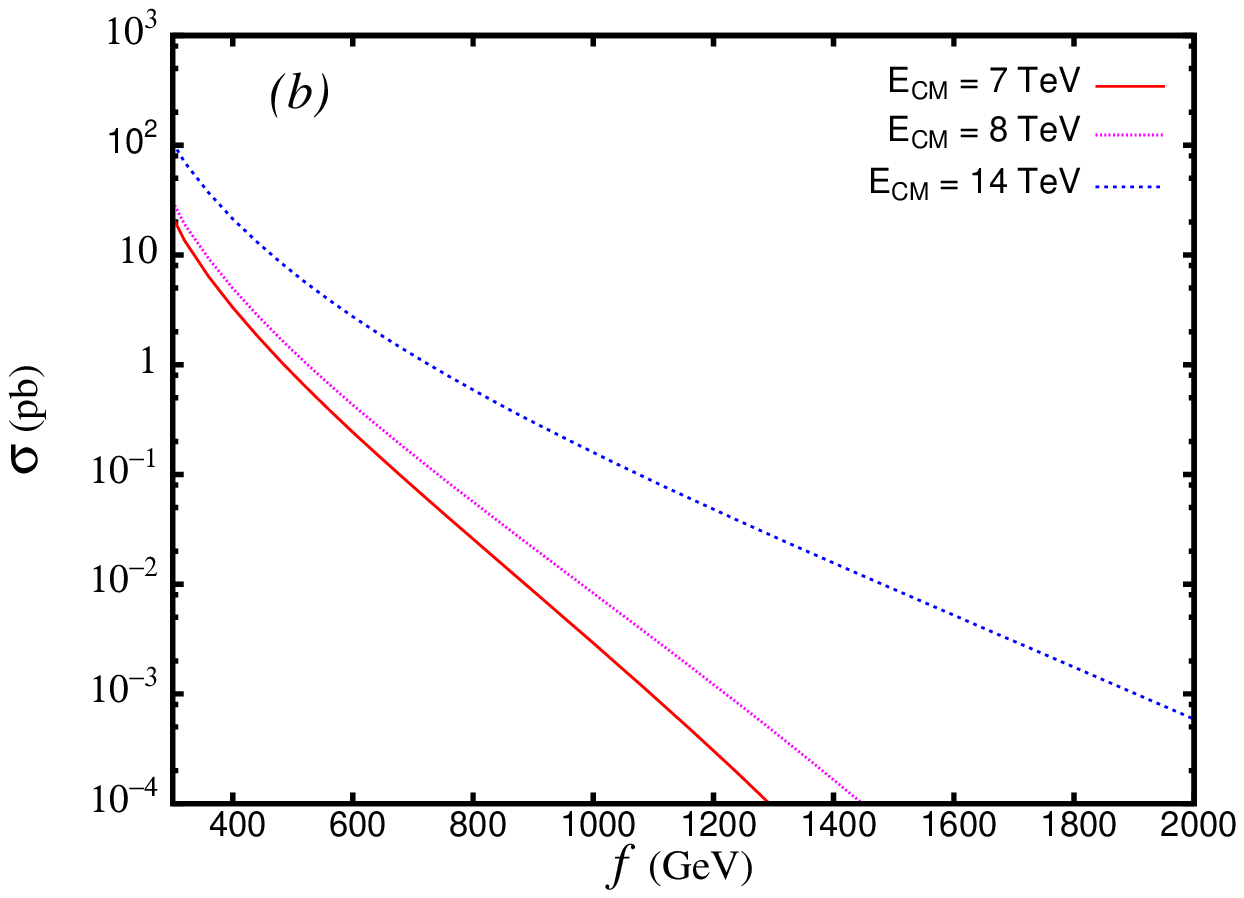}
\end{center}
\caption{\em The variation of the leading order $T$-odd quark pair $(
Q_H \bar Q_H+Q_H Q_H+\bar Q_H \bar Q_H)$ production 
with the scale $f$ for (a) $\kappa = 0.45$ and  (b) $\kappa = 1.0$ for 
the LHC running at $\sqrt{s} = 7, \, 8, \, 14 \tev$}.
\label{fig:csprod}
\end{figure}
 In our numerical analysis, we use the CTEQ6L1 parton distribution
functions \cite{cteq6l}.  In the absence of higher order calculations,
we consider only the leading-order processes.  For the pure QCD
processes, the $K$-factor, parametrizing the higher-order
contributions to the production cross section, could, in principle, be
estimated from analogous calculations for the
top-quark~\cite{K-factor} and is somewhat larger than unity. For the
electroweak processes, though, such calculations are not
available. Given this, we adopt the {\em conservative} approach of
both neglecting the $K$-factor as well as choosing a moderate
factorisation scale, viz. $Q = \sqrt{{\hat s}}/2$.  In
Fig.~\ref{fig:csprod}, we display the production rate of the $T$-odd
quark as a function of the scale $f$ for two values of the parameter
$\kappa$ namely $\kappa = 0.45 $ and $1.0$ for a few choices of the LHC operating energy.
While the pure QCD amplitude depends only on the mass of the heavy
quark, and thus on the product $\kappa f$ alone, the electroweak
amplitudes have additional dependence on $f$ (owing, {\em e.g.}, to
the $t$-channel exchange of $W_H/Z_H/A_H$ {\em etc.}). 
In addition, both the branching fractions as well as the decay distributions have
further dependence on the scale $f$.

In particular, we choose four representative values for the scale $f$ (while 
keeping fixed $\kappa = 0.45$) for all the simulations. In Table
\ref{tab:spectrum}, we list the relevant parts of the corresponding 
mass spectra as also the total production cross section for pairs 
of $Q_H$ (limiting ourselves to the partners of the first two generations). 
\begin{table}[!h]
\begin{center}
\begin{tabular}{|c|c|c|l|c|c|}\hline  
\hline \multicolumn{1}{|c|}{\multirow{2}{*}{$f$ (GeV)}} 
& \multicolumn{1}{|c|}{\multirow{2}{*}{$M_{Q_H}^{(U,D)}$ (GeV)}} 
& \multicolumn{1}{|c|}{\multirow{2}{*}{$M_{A_H}$ (GeV)}} 
& \multicolumn{3}{|c|}{$\sigma (\sum Q_H-pair)$ (fb)}
\\ \cline{4-6}
 &  &  & \multicolumn{1}{|c|}{7 TeV} 
& \multicolumn{1}{|c|}{8 TeV} 
& \multicolumn{1}{|c|}{14 TeV} \\ \hline 
\hline 750  & (470.9, 477.3) & 111.9 
& $1.62 \times 10^3$ & $2.56 \times 10^3$ & $1.42 \times 10^4$\\ \hline
\hline 1000 & (631.6, 636.4) & 153.9 
& $2.76 \times 10^2$ & $4.69 \times 10^2$ & $3.10 \times 10^3$\\ 
\hline 1500 & (951.4, 954.9) & 235.8 
& $1.51 \times 10^1$ & $3.12 \times 10^1$ & $3.26 \times 10^2$\\ 
\hline 2500 & (1589, 1591) & 397.3 
& $8.17\times 10^{-2}$ & $3.09\times 10^{-1}$ & $1.25 \times 10^1$ \\ 
\hline
\hline
\end{tabular} 
\end{center}
\caption{\em The mass spectrum for the $T$-odd quarks of the first two 
generations and the
lightest stable $T$-odd particle for four different choices of the scale $f$ 
with $\kappa=0.45$. We also list the total pair production cross section for 
the $T$-odd heavy quarks at LHC with center-of-mass energies of 7 TeV, 8 TeV
 and 14 TeV respectively.}
\label{tab:spectrum}
\end{table}

These, then, serve as our benchmark points (except for $f = 2.5 \tev$ 
at $\sqrt{s} = 7 \tev$ in view of the smallness of the cross section).

\section{Results and Analysis}
\label{sect:result_analysis}
\subsection{Dijet signal}
We now focus on the dijet plus missing energy channel in the LHT model
and compare it against the SM expectations. A hadronic machine such as
the LHC is associated with large rates for QCD processes and inclusive
dijet production is the dominant mode in SM, with cross-sections of
the order of a few {\it millibarns}. Thus, it poses the most serious
background for any new physics signal in this particular channel.
While new physics signals with strong dynamics and
resonances in the dijet invariant mass which are
expected to stand out over the QCD background have been studied in the
literature \cite{LAA,CPP,HR,BHZ,BSZ,FG,EHS,RSmodel}, no such
  resonance is expected in the LHT model.  
Instead, dijets are produced in association with the lightest massive
stable particles, which give rise to a significant amount of
missing transverse momenta. This renders difficult the observation
of such signals over the QCD background. Consequently, most search
strategies have concentrated on cascade decay modes instead. However, as 
we have already argued in the preceding section, for $\kappa \lapp 0.45$, 
the cascade decay modes disappear and the only final state available to us is
that comprising two hard jets accompanied by missing transverse momentum.

To be specific, we shall choose $\kappa = 0.45$, unless otherwise
stated.  For a given $f$, this implies the most massive $Q_H$ (and,
hence, the smallest production cross sections) consistent with a
dominant decay mode into a SM quark and the LTP.  Understandably, the
signal is completely overwhelmed by the large QCD background. Also to
be included in the background are other multijet sub-processes,
accruing dominantly from resonant processes in the SM, such as those
involving the weak gauge bosons $(W^\pm, Z)$.

The SM backgrounds, irrespective of their origin (viz, QCD, $t \bar t$
production, $W^\pm / Z$ with or without jets) were generated with {\sc
  Pythia} \cite{pythia}, thereby allowing us to include the effects of
ISR/FSR, showering and hadronization. Those backgrounds (such as
(di-)boson production with hard jets) that cannot be computed directly
thus, were generated with {\sc Alpgen}\cite{Mangano:2002ea}. These,
though, turned out to be of little consequence. The signal events, on
the other hand, were generated with CalcHEP \cite{Pukhov:2004ca} and
then interfaced with {\sc Pythia}.

The jets were constructed using the inbuilt toy calorimeter subroutine
PYCELL which is a jet clustering algorithm and is used to get the
final state jets for the analysis.
We define two jets to be separately distinguishable, if they satisfy
\beq
\Delta R_{jj} = \sqrt{(\Delta \eta)^2 + (\Delta \phi)^2} > 0.7 \ ,
   \label{eq:jet_separation}
\eeq
where $\Delta \eta$ and $\Delta\phi$ are their separations in 
rapidity and azimuthal plane respectively. Any pair of jets that does 
not satisfy this is merged. 
We require that the final state  have {\em exactly two} jets
satisfying
\begin{equation}
\begin{array}{rcl}
p_T^{j_i} &> &100 ~{\rm GeV};~(i = 1,2) \\
|\eta_j| &< &2.5  \\
\end{array}
   \label{eq:jet_requirements}
\end{equation}
where the jets ($j_1 ~\&~ j_2$) are ordered according to their
transverse momenta $(p_T^{j_i} > p_T^{j_2})$ and $\eta_j$ represents
their individual rapidities.  In other words, we veto events
containing a third jet satisfying eq.(\ref{eq:jet_requirements}).  We
also veto events with an isolated lepton ({\em i.e.}, satisfying
$\Delta R_{j\ell} > 0.4$ for each jet) with a $p_T > 10 \gev$ and
falling within the detector coverage ($|\eta_\ell| < 2.5$). To improve
the signal to background ratio, we also impose
\begin{equation}
\begin{array}{rcl}
\sum p_T^{j_1,j_2} \equiv p_T^{j_1}+p_T^{j_2}&>&500~{\rm GeV}.   \\
\end{array}
   \label{eq:basic_cuts}
\end{equation}
For future reference, we designate the combination of cuts 
in eqs.(\ref{eq:jet_requirements} \& \ref{eq:basic_cuts}) 
by ${\cal C}_{1}$. 

A final state such as ours affords very few kinematical variables 
that could be exploited to improve the signal to noise ratio. 
Indeed, the only other obvious
independent cut would be one on the missing
transverse momentum $\slashed p_T$. 
However, rather than imposing a flat requirement
on $\slashed p_T$, we instead consider a related variable $\alpha_T$ 
advocated, in an entirely different context, by Ref.\cite{Randall:2008rw}.
Expressible as the ratio of the transverse
momenta of the second leading jet and the invariant
mass of the dijet pair, it is given by
\begin{equation}
\alpha_T \equiv \frac{p_T^{j_2}}{M_{jj}} \ . 
   \label{alpha_defn}
\end{equation}
It is easy to see that, for an exclusive dijet event (i.e., one with
no other hard visible object and/or substantial $\slashed E_T$),
$p_T^{j_2} = p_T^{j_1}$ and $2 \, \alpha_T = |\sin\theta|$, with
$\theta$ being the scattering angle in the parton center of mass
frame. Thus, if $\slashed E_T$ is to originate from mismeasurements of
the jet energies, then the corresponding ratio satisfies $\alpha_T
\lapp 0.5$. Even with the emission of further jets, the situation is
not expected to vary radically as long as the extra jets are not too
hard. Indeed, as ref.\cite{Randall:2008rw} has argued, the entire QCD
background trails off beyond $\alpha_T \gapp 0.5$. The 
configuration would 
be substantially different, though, if the dijet pair
recoils against a massive 
particle (say a $Z$) or any other source of a substantial $\slashed p_T$. 
This prompts us to invoke the final selection cut (called ${\cal C}_{2}$)
namely
\begin{equation}
\alpha_T \geq  0.51 \ .
\label{eq:alpha_cut}
\end{equation}

The dominant SM background 
is that due to the  $2\to 2$ hard QCD sub-processes.  
Although the hard process, {\em per se}, is not associated 
with any missing transverse momentum, some amount of $\slashed{p}_T$ 
can arise either from the jets fragmenting into neutrinos or simply from 
a mismeasurement of the jet energy. To parametrize the latter, we effect 
a Gaussian smearing of the jet energy with a resolution given by
\[
   \frac{\Delta E}{E} = \frac{0.8}{\sqrt{E \,(\gev)}} \ . 
\]
Note that this is substantially worse than, say, the CMS resolution
in the barrel region (to which we limit our detection)~\cite{cms_hcal}, 
and, thus, represents a {\em deliberately conservative choice.}
Also included are the other backgrounds, for example, 
those emanating from  $W$+jets, $Z$+jets and $t\bar{t}$ production.

\begin{table}[!h]
\bc
\begin{tabular}{|c|c|c|c|c|c|c|c|}  \hline 
Cut &\multicolumn{3}{|c|}{Signal (fb) }&\multicolumn{4}{|c|}{}\\ \cline{2-4}
flow &\multicolumn{3}{|c|}{$f$ (TeV)}&\multicolumn{4}{|c|}{SM background (fb) }\\ \cline{2-8}
  & {0.75} & {1.0} & \multicolumn{1}{c}{1.5} & 
\multicolumn{1}{|c|}{$QCD$} &  \multicolumn{1}{|c|}{$t\bar{t}$} &  \multicolumn{1}{|c|}{$W+jets$} &  \multicolumn{1}{|c|}{$Z+jets$} \\\hline 
${\cal C}_1 $ & 84.80 & 36.10 & 5.37 & $\sim 1.62 \times 10^5$ & $92.50$ 
                                           & 710.03 & 272.02   \\
${\cal C}_1 + {\cal C}_2 $   &  2.16  & 2.79  & 1.05 & $\simeq$ 13.01 & 0.05 & 0.21 & 0.38   \\
\hline
\end{tabular}
\end{center}
\caption{\em The leading-order cross sections for the 
$2j+\slashed{E}_T$ final state at LHC with $\sqrt{s}=7 $ TeV for 
three benchmark points of the LHT model with a fixed $\kappa = 0.45 $.
We also list the rates for the dominant SM backgrounds.}
\label{tab:rates7TeV}
\end{table}

The entries in the first row of Tables \ref{tab:rates7TeV} \&
\ref{tab:rates14TeV} show the cross sections for the benchmark points
on imposition of the aforementioned selection cuts ${\cal C}_1$ (but
not ${\cal C}_2$). The suppression of the signal strength due to the
cuts is clearly discernible.  As a comparison with Table
\ref{tab:spectrum} shows, this suppression is progressively less
severe as the scale $f$ increases. This is easy to appreciate as a
larger value of $f$ implies not only larger masses for the $Q_H$, but
also larger split between $m_{Q_H}$ and $m_{A_H}$. This, in turn,
leads to harder jets from the decay of the $Q_H$, thereby satisfying
eq.(\ref{eq:basic_cuts}) with relative ease.

\begin{table}[!h]
\bc
\begin{tabular}{|c|c|c|c|c|c|c|c|c|}  \hline 
Cut  &\multicolumn{4}{|c|}{Signal (fb) }&\multicolumn{4}{|c|}{}\\ \cline{2-5}
flow &\multicolumn{4}{|c|}{$f$(TeV)}&\multicolumn{4}{|c|}{SM background (fb) }\\ \cline{2-9}
& 0.75 & 1.0 & 1.5 & 2.5 & 
\multicolumn{1}{|c|}{$QCD$} &  \multicolumn{1}{|c|}{$t\bar{t}$} &  \multicolumn{1}{|c|}{$W+jets$} &  \multicolumn{1}{|c|}{$Z+jets$} \\\hline 
${\cal C}_1 $ & $1.01 \times 10^3$ & 490.03 & 122.04 & 7.09 & $\sim 1.14 \times 10^6$ & $1.23\times 10^3 $ 
                            & $4.52\times 10^3 $ & $ 1.84\times 10^3 $ \\
${\cal C}_1 + {\cal C}_2 $   &  25.33  & 31.02  & 18.12 & 1.40 &  $\simeq$ 31.79 & 1.26 & 5.43 & 4.66   \\
\hline
\end{tabular}
\ec
\caption{\em As in Table.\ref{tab:rates7TeV}, but for $\sqrt{s} = 14 \tev$
instead.}
\label{tab:rates14TeV}
\end{table}

\begin{figure}[!ht]
\includegraphics[width=3.2in,height=3.2in]{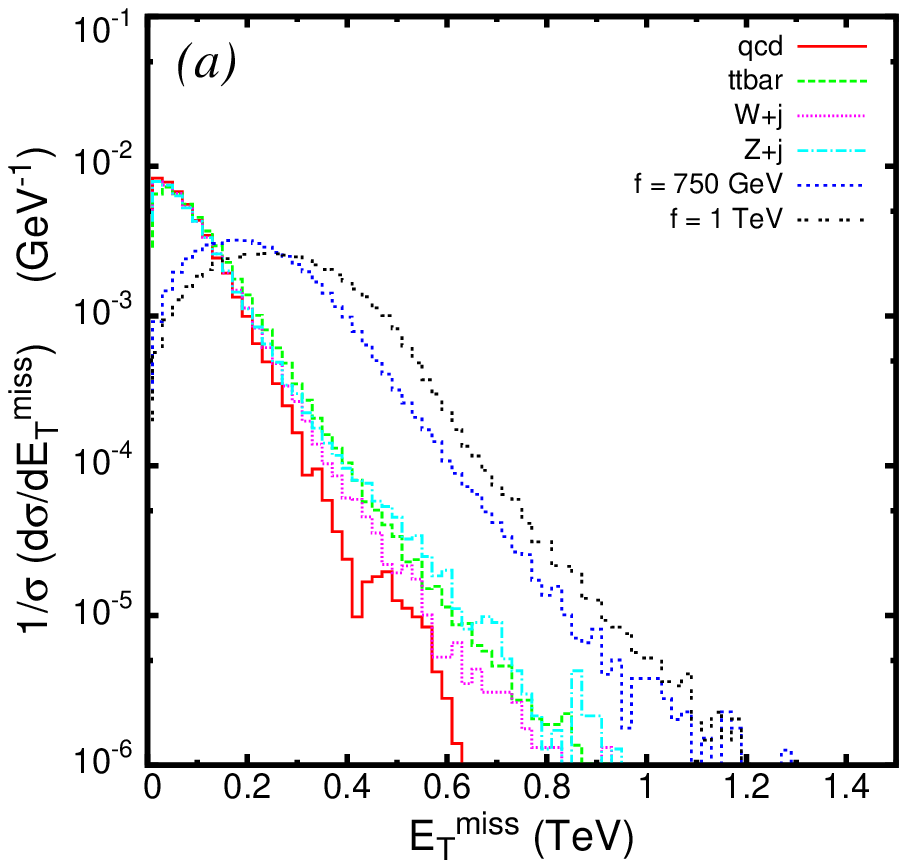}
\includegraphics[width=3.2in,height=3.2in]{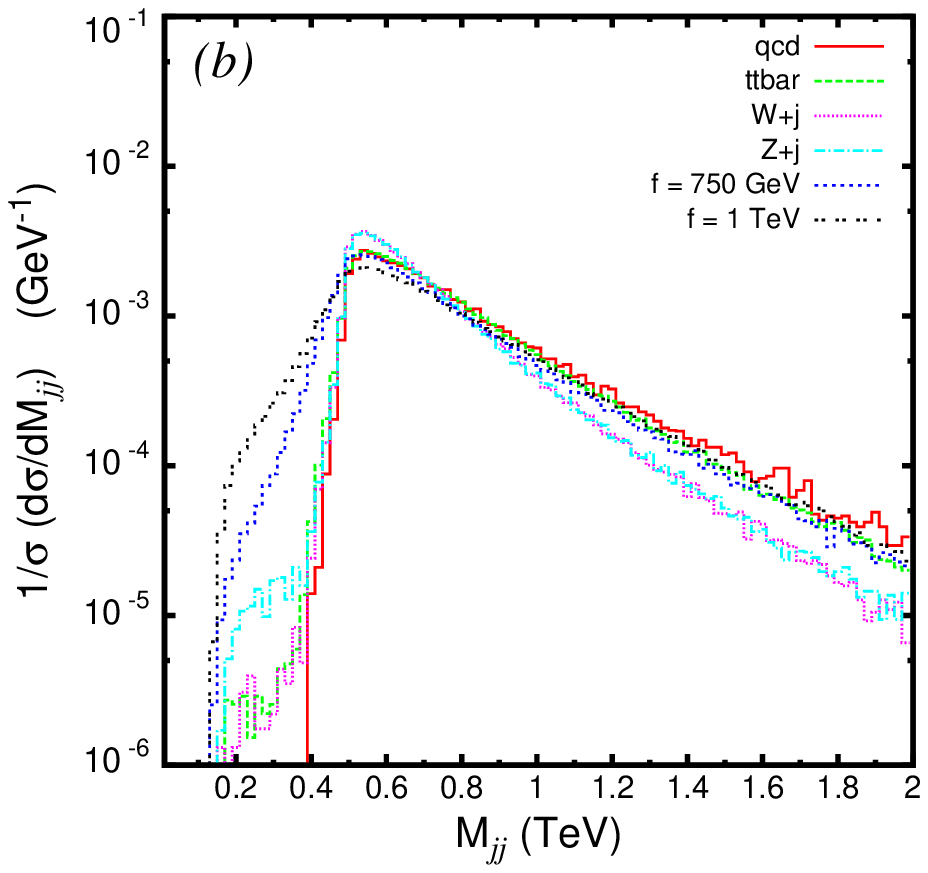}

\includegraphics[width=3.2in,height=3.2in]{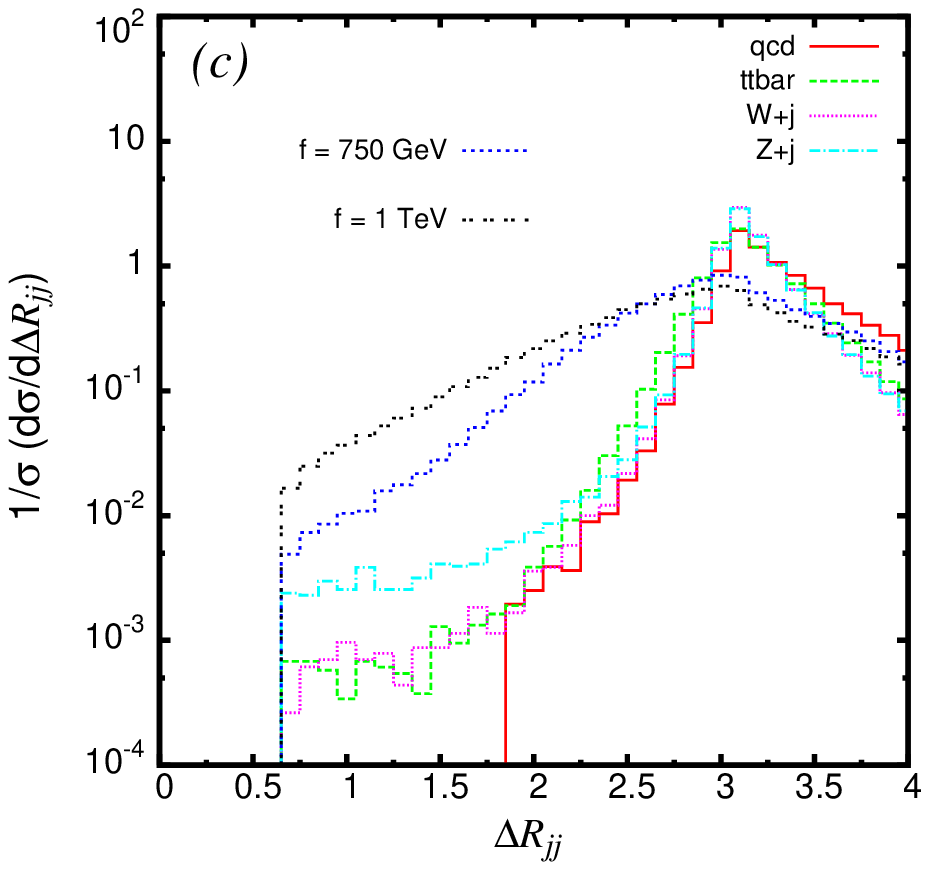}
\includegraphics[width=3.2in,height=3.2in]{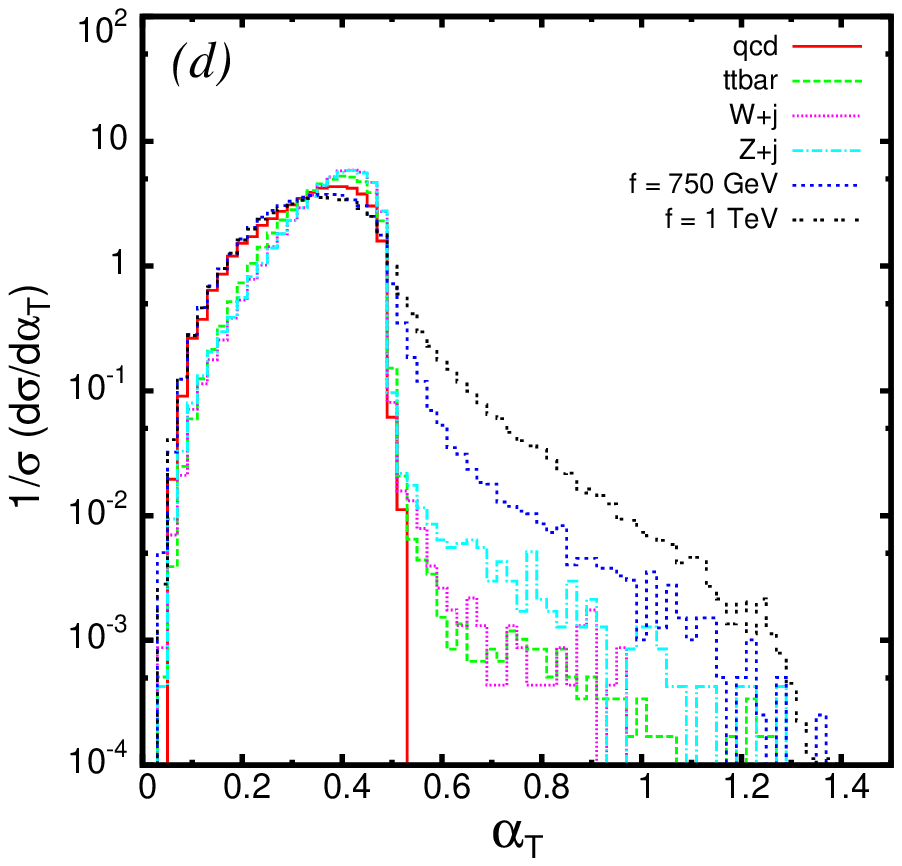}
\caption{\em Normalized differential cross sections, at the LHC
  $(\sqrt{s} = 14 \tev)$, for the signal (2 values of the scale $f$)
  and various SM background processes in {\em (a)} missing transverse
  energy, {\em (b)} the invariant mass of the two leading jets,
  $\Delta R_{jj}$ between the final state jets, {\em (c)} the cone
  angle between the same, and {\em (d)} the variable $\alpha_T$ (see
  eq.\ref{alpha_defn}).}
\label{fig:kinem}
\end{figure}

What is more important is that, on imposition of the cuts ${\cal
  C}_{1}$ alone, the QCD background is $\sim 162 \, (1140)$ pb for the
LHC operating at $\sqrt{s}=7 \, (14)$ TeV. This is orders of magnitude
larger than the signal cross sections of Tables \ref{tab:rates7TeV} \&
\ref{tab:rates14TeV}. Even the electroweak backgrounds are larger than
the signal. This necessitates the use of
additional cuts and, to this end, we must examine the phase space 
distributions. 
We present, in Fig.\ref{fig:kinem}, the normalized
distributions, for both signal and background, in various kinematical
variables\footnote{Although we present here the results for $\sqrt{s}
  = 14 \tev$, those for $\sqrt{s} = 7 \tev$ are qualitatively
  similar.}.  As Fig.\ref{fig:kinem}($a$) shows, the missing $E_T$
distribution for the background is much softer than that for the
signal. This is not unexpected as far as the QCD
background is concerned, for there the missing $E_T$ arises
largely on account of mismeasurement.  There is, of course, some
contribution from (semi-)leptonic decays of hadrons within a jet, but
these are subdominant.  The other irreducible SM contribution to this
background arises from inclusive $Z$ production followed by $Z \to
\nu_i \bar \nu_i$; the corresponding $\slashed{p}_T$ is nothing but
the transverse momentum of the $Z$ itself and, hence, is not large.
Similar is the story for inclusive $W$-production, followed by the
leptonic $W$-decays wherein the charged lepton is not registered by
the detector.

The missing $E_T$ in the
signal events, on the other hand, arises from the decay $Q_H \to q +
A_H$, with the invisible $A_H$, carrying, in the rest frame of the
$Q_H$, a momentum of $(M_{Q_H}^2 - M_{A_H}^2) / (2 \, M_{Q_H})$. For
the spectrum of Table \ref{tab:spectrum} this quantity, clearly, is
sizable. Understandably, a significant component of the transverse
momenta of the two $A_H$'s may cancel, leading to a smaller
discernible $\slashed{p}_T$. 
Even then, the spectrum would tend to be hard,
typically peaking at nearly $\slashed{p}_T \sim M_{Q_H}/2$.

The invariant mass distribution---Fig.\ref{fig:kinem}($b$)---although
being somewhat different for the signal (as compared to the background
constituents), can hardly be used efficiently to improve the signal to
noise ratio. Jet separation, on the other hand, is a very useful
variable. As Fig.\ref{fig:kinem}($c$) shows, for the bulk of the
background, the two jets are back to back. While this is readily
understandable for the QCD component, to appreciate the situation for
the rest of the SM contribution, note that this is but a restatement
of the fact that these are characterised by $\slashed{p}_T / p_T^{j_2}
\ll 1$. The signal events, on the other hand have a sizable value for
this ratio. Indeed, the distributions are correlated and imposing a
strong criterion on one would obviate doing so for the other.

However, instead of imposing cuts on $\slashed{p}_T$ or $\Delta
R_{jj}$, we rather consider the variable $\alpha_T$. As discussed
earlier, this has the advantage of correlating $\slashed{p}_T$ with
the energy scale of the event.  As promised and as demonstrated by
Fig.\ref{fig:kinem}($d$), the QCD background falls very sharply for
$\alpha_T > 0.5$. While the rates of fall for the other background
components are not as severe, the slopes are steep enough for this
variable to be considered a good signal to noise discriminator.  This
conclusion is aided by the fact that the non-QCD backgrounds were
subdominant to start with. The signal too peaks at $\alpha_T \lapp
0.5$. However, the fall beyond $\alpha_T = 0.5$ is much slower
indeed. For both the non-QCD background and signal, this extension
beyond $\alpha_T = 0.5$ owes its origin to the magnitude of
$\slashed{p}_T / p_T^{j_2}$.

It might seem at first sight that $\Delta R_{jj}$ would do the job as
well as $\alpha_T$. As Fig.\ref{fig:kinem}($c$) shows, and as we have
already discussed, the jet pair is much better separated when it comes
to the background (as compared to the signal). However, note that the
steepness of the $\Delta R_{jj}^{SM}$ distribution is not as
pronounced as that of the $\alpha_T$ distribution.  Consequently, the
improvement in the signal to background ratio is much better when we
choose to impose the cut on $\alpha_T$.  This is not surprising
because, contrary to simple kinematical variables
such as the $p_T$ of an individual jet, $\slashed{p}_T$ or $\Delta
R_{jj}$, the variable $\alpha_T$ is a correlated measure of hardness
of the event and the angular separations. This is what has prompted us
to choose cut ${\cal C}_2$ in preference to any others.

As Table \ref{tab:rates14TeV} shows, the huge background can be
effectively eliminated altogether by rejecting all events with
$\alpha_T \leq 0.51$. Whereas the majority of the signal events are
rejected as well, the signal-to-background ratio improves
dramatically.  It is interesting to note that the signal efficiency of
this $\alpha_T$ cut improves with the increase in the value of the LHT
scale $f$, although the $Q_H$ pair production cross section decreases.
This feature is easy to understand. Higher values of $f$ correspond to
a heavier LTP and T-odd quarks, which leads to larger imbalance in
energy because of heavier LTP which in turn causes a wider spread in
the $\alpha_T$ distribution (see Fig. \ref{fig:kinem} ($d$)) for $f=1$
TeV when compared to $f=750$ GeV. As a result of this, one expects
that the $\alpha_T$ cut is less severe for higher values of $f$
leading to better signal significance. Thus, we expect that even
though the pair production cross section for T-odd quarks may decrease
for large $f$, the kinematic selection that is most effective in
suppressing the SM background also makes the signal significance
better for large values of $f$. We must, however, note that eventually
the small production cross section for very large values of $f$ will
take over and make the signal too small to be significant.

The corresponding distributions---for both the signal and
background---look very similar for LHC operating at $\sqrt{s}=7$ TeV,
although the event rates are much smaller.  Consequently, the
$\alpha_T$ cut would work as well for that case.  Based on the
preceding analysis, a quick and naive estimate of the LHC sensitivity
can be made by simply observing the strength of the signal and
background events shown in Tables \ref{tab:rates7TeV} \&
\ref{tab:rates14TeV}. For the LHC at $\sqrt{s} = 7 $ TeV and with the
current integrated luminosity ${\cal L} \sim 5 ~{\rm fb}^{-1}$ the
sensitivity $(N_s /\sqrt{N_s+ N_b})$ is less than $2\sigma $ for all
the three benchmark points.

It has been now announced that the LHC
will run at $\sqrt{s} = 8 $ TeV during 2012 and is also expected to
deliver a luminosity of $15 {\rm fb}^{-1}$ for both CMS and
ATLAS. Consequently, the total heavy $T$-odd quark pair production
cross-section would increase by a factor varying between 1.16 - 2 for the LHT 
scale $f = 0.75 -1.5 $TeV. Even accounting for the increase in the background, 
the two upgrades, together, would imply that each of ATLAS/CMS would be 
in a position to report a significant excess by the end of the year if 
$f \lapp 1.5 \tev$.
A future upgrade of the LHC center of mass
energy to $\sqrt{s} = 14 $ TeV will be able to probe the LHT model
with $f = 0.75 - 1.5 $ TeV at $5\sigma $ significance in the
$2j+\slashed{E}_T$ channel with an integrated luminosity as less as
${\cal L } = 3-5~{\rm fb}^{-1}$.

\subsection{Heavy flavor-tagged dijet signal}
\begin{figure}[!h]
\begin{center}
\includegraphics[width=3.5in]{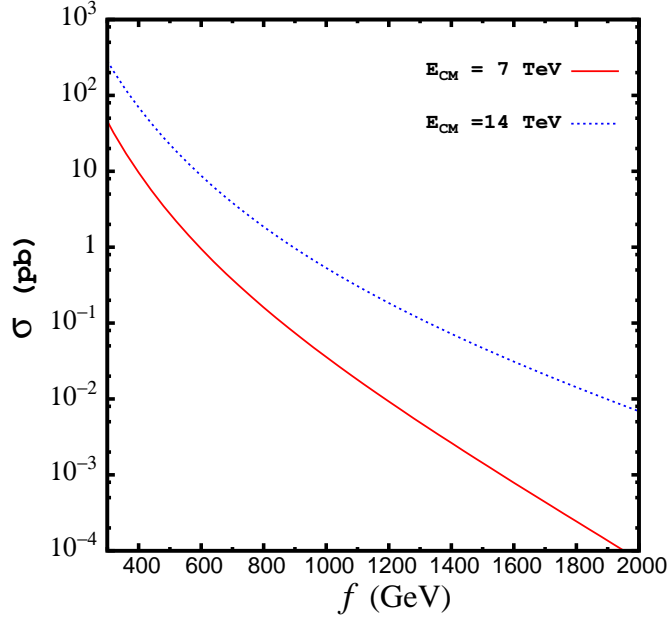}
\end{center}
\caption{\em The variation of the leading order $T$-odd quark pair 
$B_H B_H, B_H \bar B_H$ production with the scale $f$ for $\kappa = 0.45$  
with the 
LHC running at $\sqrt{s} = 7 $ TeV and 14 TeV.}
\label{fig:csBB}
\end{figure}

We now specialize to the case where both the jets are
$b$-tagged. Recent analyses at both ATLAS and CMS have shown that a
very high efficiency for $b$-tagging may be
obtained~\cite{ATLAS_btag,CMS_btag}.  Dependent on the transverse
momenta of the $b$-jets, the efficiencies are as high as 70\% for jets
with $p_T > 100$ GeV.  We work with the same representative
  points shown in Table \ref{tab:spectrum}, and where the $T$-odd
  $B_H$ decays to the LTP and $b$-quark with 100\% probability for
  $\kappa < 0.45$, as shown in Fig.\ref{fig:brfig}. Since the mass of
  $B_H$ is degenerate with $D_H$ one expects similar production cross
  section for them to be pair produced as the $D_H$ states except that
  the only significant contribution to the cross section comes from
  the QCD dominated sub-processes.  In Fig.~\ref{fig:csBB} we plot
the cross section for the process $pp \to B_H B_H (\bar B_H) $
  at LHC.

We note that the pair production cross section is about an order of magnitude 
smaller when compared to the pair production cross section 
of the first two generation of the $T$-odd heavy quarks. However, one expects 
much smaller background for the signal, once the final state
jets are tagged as $b$-jets. The dominant background with very little missing 
transverse momenta would be the QCD production of $b\bar b+X$. In addition the 
other dominant background comes from the QCD dijet sub-processes for lighter 
quarks, with the light flavor jets mistagged as $b$-jets. Though the mistag efficiency for such high $p_T$ light jets is less than 1\%, 
the sheer enormity of the QCD cross section in SM makes 
it a serious background for the signal. The other major 
SM backgrounds are due to $W+jets, Z+jets$ and $t\bar t$. Note that guided by 
the previous analysis for the dijet signal, one can easily repeat the same 
requirements on the phase space to suppress the SM background. For our analysis
 we demand that both the 
final state high $p_T$ jets are tagged as $b$-jets. By doing this we do have a 
significant suppression for the signal, but demanding such hard $b$-jets with 
the kinematic cuts mentioned in Eqs. (\ref{eq:basic_cuts}) and 
(\ref{eq:alpha_cut}) effectively suppresses almost all of the SM background, including that 
from $W+b\bar{b}+X$ and $t\bar{t}$  ($\lesssim 10^{-3}$ fb).

\begin{table}[!t]
\begin{center}
\begin{tabular}{|c|c|c|c|} \hline
 \multicolumn{3}{|c|}{Signal (fb) } & SM background (fb)\\ \cline{1-3}
  $f=750$ GeV & $f=1$ TeV & $f=1.5$ TeV & \\ 
\hline  0.523 (0.045) &  0.767 (0.059) & 0.465 (0.023) & $\simeq$ 0.242 (0.014) \\ 
\hline 
\end{tabular} 
\end{center}
\caption{\em The leading-order cross sections for the $2b-jets +\slashed{E}_T$ 
final state at LHC with 
$\sqrt{s}=14 (7)$ TeV for the LHT model with three different choices of the scale $f$ 
and $\kappa=0.45$.The cuts are the same as ${\cal C}_1+{\cal C}_2$ except for $\Delta R_{bb}>0.5$}
\label{tab:ratesbb}
\end{table}
For identifying $b$-jets with the most energetic transverse jets
selected we put the condition: a jet is tagged as $b$-jet if it is
associated to a parent $b$-quark. The identification is made by
demanding that the most energetic parton within a cone of radius
$\Delta R=0.5$ around the jet axis is a $b$ quark and also that the
opening angle between the $b$ quark and the jet axis lies within
$20^0$.  As our $b$-jets have $p_T>100$ GeV, we assume the average
$b$-tagging efficiency of 50\%. For light jets (jets originating in
$u,d,c,s$ quarks or in gluons), a fake $b$-tagging efficiency of 1\%
is assumed We consider the same set of kinematics cuts as before
except that we put $\Delta R_{bb} > 0.5$. We find that the strong cuts
on the $b$-jet $p_T$ and on the scalar sum along with the $\alpha_T$
cut makes the SM background significantly small albeit comparable to
the signal. We list the signal and background cross sections (after
all cuts and multiplying with the respective efficiencies) in Table
\ref{tab:ratesbb}. Remembering that what we are looking here is a
small subset of the dijet process with reduced efficiencies, it is
quite natural to find a much suppressed rate for the $2b$-jets
$+\slashed{E}_T$ signal as compared to the dijet $+\slashed{E}_T$
final state shown in Tables \ref{tab:rates7TeV} \&
\ref{tab:rates14TeV}.  Understandably, with the present luminosity
$(\sim 5 {\rm fb}^{-1}) $, we do not expect any significant excess
over the SM prediction for this particular channel. We do not expect
any significant enhancement in the $N_s/\sqrt{N_s+N_b}$ even with an
increase of center of mass energy of the LHC to 8 TeV with $15 {\rm
  fb}^{-1}$ data.  However, $3\sigma $ signal sensitivity can be
obtained for the LHT scale $f =1 $ TeV with an integrated luminosity
in excess of $10$ fb$^{-1}$ at LHC with $\sqrt{s}=14$ TeV center of
mass energy.

\section{Conclusions}
\label{sect:concls}
A phenomenological ``imperative'' in Little Higgs models is the
introduction of a $T$-parity. Apart from predicting a candidate for
cold dark matter, this also leads to the presence, in the spectrum, of
relatively light $T$-odd quarks, $Q_{iH}$ which can be copiously
pair-produced at the LHC.  Canonical search strategies for the same
have concentrated on the cascade decay of the $Q_{iH}$ through the
$T$-odd counterparts of the $Z$ and $W$. For a significantly large
fraction of the Littlest Higgs model parameter space, though, such
cascades are kinematically forbidden and the $Q_{iH}$ decay directly
to a single SM quark and the dark matter candidate $A_H$.

In this paper, we investigate this very decay, which leads to a final
state comprising of a dijet pair alongwith a large missing transverse
momentum. To this end, we simulate the production of all possible
pairs of $Q_{iH} \, \bar Q_{jH} \, (Q_{jH})$, including the
electroweak processes.  We perform both a parton-level Monte-Carlo,
and then generate events using CalcHEP~\cite{Pukhov:2004ca} interfaced
with {\sc Pythia} \cite{pythia} thereby allowing us to include the
effects of ISR/FSR, showering and hadronization. The jets are
constructed using the standard {\sc Pythia} routine PYCELL.  The major
standard model background for this signal comes from the pure QCD
dijet events (with the missing $p_T$ accruing from both the
mis-measurement of jet energies as well as the fragmentation into
neutrinos), with $W+jets, Z+jets $ and $t {\bar t}$ processes also
contributing handsomely.  All these too are generated using {\sc
  Pythia}, and cross-checked with {\sc Alpgen}~\cite{Mangano:2002ea}.

While requiring that there be only two jets with rather stringent
demands the scale of the hadronic activity does serve to substantially
reduce the background, the latter still overwhelms the signal
size. Further demands on the magnitude of missing-$E_T$ does improve
the signal-to-background ratio, but it is still not enough (owing
largely to the fact that the large hadronic activity itself results in
a significant $E_T^{\rm miss}$ accruing from energy mismeasurements).
Jet angular separation, on the other hand, plays a crucial role.
Although $\Delta R_{jj}$ can be used profitably, we find that
correlating the jet $p_T$ and their separation through the
introduction of the $\alpha_T$ variable (see eq.\ref{alpha_defn}) is
a far superior alternative.  Imposing $\alpha_T > 0.51$ almost
entirely eliminates the dominant QCD background, and also reduces the
other significantly. 

We observe though that, for the LHC at $\sqrt{s} = 7 $ TeV and with 
the current integrated luminosity
${\cal L} \sim 5 ~{\rm fb}^{-1}$, the statistical significance $(N_s /\sqrt{N_s+ N_b})$ 
would be less than $2\sigma $ for all the three benchmark points. 
However, the recently announced upgrades to a
center of mass energy of 8 TeV and an integrated luminosity of 
$15~{\rm fb}^{-1}$ during 2012 raises tantalizing propspects of discovery 
within the year. Finally, if the LHC attains 
the center of mass energy of $\sqrt{s} = 14 $ TeV it 
will be able to probe the LHT model with $f = 0.75 - 1.5 $ TeV at $5\sigma $ 
significance in the $2j+\slashed{E}_T$ channel with an integrated luminosity 
as less as  ${\cal L } = 3-5~{\rm fb}^{-1}$. 

On the other hand, the pair production of heavy $T$-odd quark $B_H$
lead to the $2b-jets +\slashed{E}_T $ signal, where we tag both the
$b$-jets. We find that due to suppressed signal cross-section, probing
the LHT model via this particular channel is in fact almost impossible
for the LHC operating at $\sqrt{s} = 7 $ TeV and the situation remains
unchanged even at $\sqrt{s} = 8 $ TeV. However, at 14 TeV LHC, with an
integrated luminosity of $\sim 20~{\rm fb}^{-1}$, we expect to
independently probe (with a $3\sigma $ significance) the the
LHT model in this channel up to a scale $f =1 $ TeV.

Before we conclude, it is worth mentioning that the both the
ATLAS~\cite{Aad:2011cw,Aad:2011ib} and
CMS~\cite{Chatrchyan:2011zy,Collaboration:2011ida,Chatrchyan:2011bj}
collaborations have analysed jets plus missing transverse momentum
signal in the context of supersymmetric scenarios.  However, it should
be noted that the difference in spectra between the two cases leads to
a marked difference in the cut efficiencies.  In other words, the LHT
parameter space discussed by us remains unconstrained by the present
ATLAS/CMS analyses.  Consequently, in our analysis, we have advocated
the use of an alternate set of selection cuts, which would serve to
increase the sensitivity.

\section*{Acknowledgements}
D.C. and D.K.G. thank ICTP High Energy Group for their hospitality
during a phase of the work.  D.K.G. also acknowledges partial support
from the Department of Science and Technology, India under the grant
SR/S2/HEP-12/2006. S.K.R. would like to thank A. Khanov for letting
him use the cluster facility. S.K.R. is supported in part by the US
Department of Energy, Grant Number DE-FG02-04ER41306.


\begin{thebibliography}{99}

\bibitem{lh0}N.~Arkani-Hamed, A.~G.~Cohen and H.~Georgi, Phys. \ Lett. \ B 
{\bf 513}, 232 (2001); \\
For reviews, see, for example, \\
M.~Schmaltz and 
D.~Tucker-Smith, Ann.\ Rev.\ Nucl.\ Part.\ Sci.\ {\bf 55},229 (2005); \\
M.~Perelstein, Prog.\ Part.\ Nucl.\ Phys.\ {\bf 58}, 
247 (2007), and references therein.

\bibitem{lh_ew}
  C.~Csaki, J.~Hubisz, G.~D.~Kribs, P.~Meade and J.~Terning,
  Phys.\ Rev.\ D {\bf 67}, 115002 (2003);\\
  J.~L.~Hewett, F.~J.~Petriello and T.~G.~Rizzo,
  JHEP {\bf 0310}, 062 (2003);\\
  C.~Csaki, J.~Hubisz, G.~D.~Kribs, P.~Meade and J.~Terning,
  Phys.\ Rev.\ D {\bf 68}, 035009 (2003);\\
  M.~C.~Chen and S.~Dawson,
  Phys.\ Rev.\ D {\bf 70}, 015003 (2004);\\
  W.~Kilian and J.~Reuter,
  Phys.\ Rev.\ D {\bf 70}, 015004 (2004);\\
  Z.~Han and W.~Skiba,
  Phys.\ Rev.\ D {\bf 71}, 075009 (2005).
  \bibitem{lht1}
  I.~Low,
  JHEP {\bf 0410}, 067 (2004).
\bibitem{lht2}
  J.~Hubisz and P.~Meade,
  Phys.\ Rev.\ D {\bf 71}, 035016 (2005).

\bibitem{lht3}
  J.~Hubisz, P.~Meade, A.~Noble and M.~Perelstein,
  JHEP {\bf 0601}, 135 (2006).

\bibitem{lht4}
  H.~C.~Cheng and I.~Low,
  JHEP {\bf 0309}, 051 (2003);
  JHEP {\bf 0408}, 061 (2004).
\bibitem{wyler}
A.~Freitas and D.~Wyler, JHEP {\bf 0611}, 061 (2006).

\bibitem{Belyaev:2006jh}
  A.~Belyaev, C.~R.~Chen, K.~Tobe and C.~P.~Yuan,
  Phys.\ Rev.\  D {\bf 74}, 115020 (2006).
\bibitem{Carena:2006jx}
  M.~S.~Carena, J.~Hubisz, M.~Perelstein and P.~Verdier,
  Phys.\ Rev.\  D {\bf 75}, 091701 (2007).
\bibitem{Chen:2006ie}
  C.~S.~Chen, K.~Cheung and T.~C.~Yuan,
  Phys.\ Lett.\  B {\bf 644}, 158 (2007).
\bibitem{Choudhury:2006mp}
  D.~Choudhury and D.~K.~Ghosh,
  JHEP {\bf 0708}, 084 (2007).

\bibitem{Cacciapaglia:2009cu}
  G.~Cacciapaglia, S.~R.~Choudhury, A.~Deandrea, N.~Gaur and M.~Klasen,
  JHEP {\bf 1003}, 059 (2010).

\bibitem{Cacciapaglia:2009cv}
  G.~Cacciapaglia, A.~Deandrea, S.~R.~Choudhury and N.~Gaur,
  Phys.\ Rev.\  D {\bf 81}, 075005 (2010).
  
\bibitem{lht_dm}
  M.~Asano, S.~Matsumoto, N.~Okada and Y.~Okada,
  arXiv:hep-ph/0602157;\\
  A.~Birkedal, A.~Noble, M.~Perelstein and A.~Spray,
  Phys.\ Rev.\ D {\bf 74}, 035002 (2006).

\bibitem{Bhattacherjee:2009jh}
  B.~Bhattacherjee, A.~Kundu, S.~K.~Rai and S.~Raychaudhuri,
  Phys.\ Rev.\  D {\bf 81}, 035021 (2010).

\bibitem{Combridge:1978kx}
  B.~L.~Combridge,
  Nucl.\ Phys.\  B {\bf 151} (1979) 429.
  

\bibitem{ArkaniHamed:2004yi}
  N.~Arkani-Hamed, S.~Dimopoulos, G.~F.~Giudice and A.~Romanino,
  Nucl.\ Phys.\  B {\bf 709}, 3 (2005)
  [arXiv:hep-ph/0409232];\\
  D.~Choudhury, S.~K.~Gupta and B.~Mukhopadhyaya,
  Phys.\ Rev.\  D {\bf 78}, 015023 (2008)
  [arXiv:0804.3560 [hep-ph]];\\
  S.~Bressler  [ATLAS Collaboration and CMS Collaboration],
  arXiv:0710.2111 [hep-ex].


\bibitem{Perelstein:2011ds}
  M.~Perelstein and J.~Shao,
  Phys.\ Lett.\  B {\bf 704}, 510 (2011)
  [arXiv:1103.3014 [hep-ph]].
  
\bibitem{cteq6l}
  J.~Pumplin, A.~Belyaev, J.~Huston, D.~Stump and W.~K.~Tung,
  JHEP {\bf 0602}, 032 (2006)
  [arXiv:hep-ph/0512167].

\bibitem{K-factor}
  M.~Cacciari, S.~Frixione, M.~L.~Mangano, P.~Nason and G.~Ridolfi,
  JHEP {\bf 0809}, 127 (2008)
  [arXiv:0804.2800 [hep-ph]];\\
\textit{See also,}
  S.~Moch and P.~Uwer,
  Phys.\ Rev.\  D {\bf 78}, 034003 (2008)
  [arXiv:0804.1476 [hep-ph]]; \\
  N.~Kidonakis and R.~Vogt,
  Phys.\ Rev.\  D {\bf 78}, 074005 (2008)
  [arXiv:0805.3844 [hep-ph]].

\bibitem{LAA} L.~A.~Anchordoqui et al., Phys.\ Rev.\ Lett. {\bf 101}, 241803 
(2008).

\bibitem{CPP} S.~Cullen, M.~Perelstein and M.~E.~Peskin, 
Phys.\ Rev.\ D {\bf 62},055012 (2000).


\bibitem{HR} J.~L.~Hewett and T.G. Rizzo, Phys.\ Rept. {\bf 183},193 (1989).


\bibitem{BHZ}U.~Baur,I.~Hinchliffe and D.~Zeppenfeld, 
Int.\ J.\ Mod.\ Phys.\ A {\bf 2}, 1285 (1987).


\bibitem{BSZ}U.~Baur, M.~Spira and P.~M.~Zerwas, 
Phys.\ Rev.\ D {\bf 42},815 (1990).

\bibitem{FG}P.~H.~Frampton and S.~L.~Glashow, 
Phys.\ Lett.\ B {\bf 190},157 (1987)

\bibitem{EHS} E.~H.~Simmons, Phys.\ Rev.\ D {\bf 55}, 1678 (1997).
\bibitem{RSmodel}L.~Randall and R.~Sundrum, Phys.\ Rev.\ Lett. {\bf 83},
4690 (1999).
  
\bibitem{pythia}
  T.~Sjostrand, S.~Mrenna and P.~Z.~Skands,
  JHEP {\bf 0605}, 026 (2006)
  [arXiv:hep-ph/0603175].
  
 \bibitem{Mangano:2002ea}
  M.~L.~Mangano, M.~Moretti, F.~Piccinini, R.~Pittau and A.~D.~Polosa,
  JHEP {\bf 0307}, 001 (2003)
\bibitem{Pukhov:2004ca}
  A.~Pukhov,
  arXiv:hep-ph/0412191.

\bibitem{Randall:2008rw}
  L.~Randall and D.~Tucker-Smith,
  Phys.\ Rev.\ Lett.\  {\bf 101}, 221803 (2008)

\bibitem{cms_hcal}P.~Schieferdcker et al., CMS Analysis Note-2008/001 (2008);\\
   CMS Collaboration, {\em CMS HCAL Technical Design Report}, 
CERN/LHCC 97-31 (1997).

\bibitem{ATLAS_btag} ATLAS Collaboration, 
Report numbes : ATLAS-CONF-2011-102;
ATLAS-CONF-2011-089;
 
\bibitem{CMS_btag} CMS Collaboration, 
Report Number: CMS-PAS-BTV-11-001.

\bibitem{Aad:2011cw} 
  [ATLAS Collaboration],
  arXiv:1112.3832 [hep-ex].

\bibitem{Aad:2011ib} 
  G.~Aad {\it et al.}  [ATLAS Collaboration],
  arXiv:1109.6572 [hep-ex].

\bibitem{Chatrchyan:2011zy} 
  S.~Chatrchyan {\it et al.}  [CMS Collaboration],
  Phys.\ Rev.\ Lett.\  {\bf 107}, 221804 (2011)
  [arXiv:1109.2352 [hep-ex]].
\bibitem{Collaboration:2011ida} 
  S.~Chatrchyan {\it et al.}  [CMS Collaboration],
  JHEP {\bf 1108}, 155 (2011)
  [arXiv:1106.4503 [hep-ex]].
\bibitem{Chatrchyan:2011bj} 
  S.~Chatrchyan {\it et al.}  [CMS Collaboration],
  JHEP {\bf 1107}, 113 (2011)
  [arXiv:1106.3272 [hep-ex]].

\end{thebibliography}
\end{document}